\documentclass[twocolumn,english,superscriptaddress,fleqn,10pt,aps,pra]{revtex4-1}%,manuscript
\usepackage[T1]{fontenc}
\usepackage[utf8]{luainputenc}
\usepackage{graphicx}
\usepackage{subfigure}
\usepackage{amsmath}
\usepackage{amsfonts}
\usepackage{wasysym}
\usepackage{mathtools}

\usepackage{parskip}
\usepackage[colorlinks,
linkcolor=black,
anchorcolor=blue,
citecolor=blue,
urlcolor=blue
]{hyperref}

\makeatletter
\@ifundefined{textcolor}{}
{
\definecolor{BLACK}{gray}{0}
\definecolor{WHITE}{gray}{1}
\definecolor{RED}{rgb}{1,0,0}
\definecolor{GREEN}{rgb}{0,1,0}
\definecolor{BLUE}{rgb}{0,0,1}
\definecolor{CYAN}{cmyk}{1,0,0,0}
\definecolor{MAGENTA}{cmyk}{0,1,0,0}
\definecolor{YELLOW}{cmyk}{0,0,1,0}
}

\makeatother

\usepackage{babel}
\begin{document}

\title{Magnetic-enhanced modulation transfer spectroscopy and laser locking for $^{87}\text{Rb}$ repump transition}

\author{Jin-Bao Long}
\affiliation{Hefei National Laboratory for Physical Sciences at Microscale and Department of Modern Physics, University of Science and Technology of China, Hefei, Anhui 230026, China}
\affiliation{CAS Center for Excellence and Synergetic Innovation Center in Quantum Information and Quantum Physics, University of Science and Technology of China, Hefei, Anhui 230026, China}
\affiliation{CAS-Alibaba Quantum Computing Laboratory, Shanghai 201315, China}

\author{Sheng-Jun Yang}
\email{yangsj85@ustc.edu.cn}
\affiliation{Hefei National Laboratory for Physical Sciences at Microscale and Department of Modern Physics, University of Science and Technology of China, Hefei, Anhui 230026, China}
\affiliation{CAS Center for Excellence and Synergetic Innovation Center in Quantum Information and Quantum Physics, University of Science and Technology of China, Hefei, Anhui 230026, China}
\affiliation{CAS-Alibaba Quantum Computing Laboratory, Shanghai 201315, China}

\author{Shuai Chen}
\email{shuai@ustc.edu.cn}
\affiliation{Hefei National Laboratory for Physical Sciences at Microscale and Department of Modern Physics, University of Science and Technology of China, Hefei, Anhui 230026, China}
\affiliation{CAS Center for Excellence and Synergetic Innovation Center in Quantum Information and Quantum Physics, University of Science and Technology of China, Hefei, Anhui 230026, China}
\affiliation{CAS-Alibaba Quantum Computing Laboratory, Shanghai 201315, China}

\author{Jian-Wei Pan}
\affiliation{Hefei National Laboratory for Physical Sciences at Microscale and Department of Modern Physics, University of Science and Technology of China, Hefei, Anhui 230026, China}
\affiliation{CAS Center for Excellence and Synergetic Innovation Center in Quantum Information and Quantum Physics, University of Science and Technology of China, Hefei, Anhui 230026, China}
\affiliation{CAS-Alibaba Quantum Computing Laboratory, Shanghai 201315, China}

%%%%%%%%%%%%%%%%%%% abstract %%%%%%%%%%%%%%%%

\begin{abstract}
Locking of a laser frequency to an atomic or molecular resonance line is a key technique in applications of laser spectroscopy and atomic metrology. Modulation transfer spectroscopy (MTS) provides an accurate and stable laser locking method which has been widely used. Normally, the frequency of the MTS signal would drift due to Zeeman shift of the atomic levels and rigorous shielding of stray magnetic field around the vapor cell is required for the accuracy and stability of laser locking. Here on the contrary, by applying a transverse bias magnetic field, we report for the first time observation of a magnetic-enhanced MTS signal on the transition of $^{87}\text{Rb}$ $D_2$-line $F_g\!=\!1\!\to\!F_e\!=\!0$ (close to the repump transition of $F_g\!=\!1\!\to\!F_e\!=\!2$), with signal to noise ratio larger than 100:1. The error signal is immune to the external magnetic fluctuation. Compared to the ordinary MTS scheme, it provides a robust and accurate laser locking approach with more stable long-term performance. This technique can be conveniently applied in areas of laser frequency stabilization, laser manipulation of atoms and precision measurement.
\end{abstract}

\maketitle
%%%%%%%%%%%%%%%%%%%%%%%%%%  body  %%%%%%%%%%%%%%%%%%%%%%%%%%
\section{Introduction}
\label{intro}

Laser frequency stabilization is critical in applications of precision spectroscopy and atom physics~\cite{Demtroder2003, Cronin2009, Ludlow2015}. Various kinds of spectroscopy techniques are applied to stabilize the laser frequency on certain atomic or molecular transition lines, including saturated absorption spectroscopy (SAS)~\cite{Schmidt1994, Liu2012, Wang2015}, dichroic atomic vapor laser locking (DAVLL)~\cite{Corwin1998, Wasik2002, Harris2008, Su2014}, polarization spectroscopy~\cite{Wieman1976, Harris2006, Do2008, Kunz2013, Torrance2016}, frequency modulation spectroscopy (FMS)~\cite{Bjorklund1980, Ben-Aroya2008, Eichholz2013}, modulation transfer spectroscopy (MTS)~\cite{Shirley1982, Schenzle1982, Jaatinen1995, McCarron2008}, and hybrid method combing the FMS and MTS~\cite{Zi2017}, etc. Besides, for more flexible frequency stabilization, alternative approach is to use a cavity~\cite{Black2001, Millo2009} or a wavelength meter~\cite{Kobtsev2007, Metzger2017, Couturier2018} for reference. Among all these methods, the MTS performs appreciable linewidth narrowing and stability, and is well suited for laser locking onto a well-defined atom frequency reference. Since the initial papers on the MTS technique~\cite{Shirley1982, Schenzle1982}, it has been fully studied and applied for over three decades~\cite{Bertinetto2001, Zhang2003, McCarron2008, Jaatinen2009, Noh2011, Negnevitsky2013, Sun2016}.

The MTS is usually a nonlinear four-wave mixing (FWM) phenomenon~\cite{Shirley1982, Berman2008}, where a modulated pump beam and a probe beam are counter-propagating. Interacted with the non-linear atom media, the sideband for the probe field, as the forth signal, is generated due to the probe beam and two frequency components of the modulated pump beams. The MTS signals are observed by means of detecting the beat signal between the probe field and the induced sidebands of it with a phase sensitive detector. Significant signals are obtained on the cycling transitions where atoms are not dissipated to the other states~\cite{Li2011}. For the other non-cycling transitions, the signals are usually negligible. Thus, the generated dispersive-like line shapes sit on a flat zero background and are insensitive to the background absorption of the media. Because the MTS only takes place when the sub-Doppler resonance condition is satisfied, the sub-Doppler resolution and steep signal gradient across the resonant point can be achieved.

However, in the ordinary MTS scheme, stray magnetic field would induce unwanted energy shift of the Zeeman sub-levels and fluctuation of atom state distribution. The reference atom line is unstable. The MTS signal would also be distorted and weakened, which would affect the stability and accuracy of the laser locking. So, the atom vapor cell needs elaborately $\mu$-metal magnet-shielding. In this work, we report the experimental observation of a magnetic-enhanced MTS signal on the transition of $^{87}\text{Rb}$ $D_2$ line $F_g\!=\!1\!\to\!F_e\!=\!0$, when a bias magnetic field with several Gauss is applied across the vapor cell. The bias field significantly improves the signal amplitude for the $F_g\!=\!1\!\to\!F_e\!=\!0$ transition, and makes it comparable with that of the $F_g\!=\!2\!\to\!F_e\!=\!3$ transition. The signals from the other closely hyperfine levels are also effectively suppressed. The signal to noise ratio, that comparison between the peak-to-peak signal amplitude and the fluctuation noise after laser lock, reaches $100\!:\!1$ or more. On the contrary, in the ordinary MTS scheme, the signal on this transition is utterly overwhelmed in the background noise. Due to the magnetic-insensitive states of $m_{F_{g,e}}\!=\!0$ in our scheme, the resonant point has little drift and immune from the magnetic fluctuations. Stray magnetic shielding around the vapor cell is not harsh. The laser can be easily locked on the enhanced MTS signal and reach a promising long-term frequency stability. This transition is close to the repump transition of $F_g\!=\!1\!\to\!F_e\!=\!2$, providing a flexible setup choice in lots of experiment research, especially for laser cooling and Raman spectroscopy.

In this paper, we present our experiment results on the magnetic-enhanced MTS spectroscopy and laser locking method. Theoretical description of this method is presented and various influences of these signals are analyzed. The structure of the paper is as follows. Section~\ref{sec:theory} outlines physical mechanism of the MTS process, particularly considering Zeeman splitting of the atomic energy levels. In Section~\ref{sec:setup}, we describe the experimental setup. Then, we illustrate the main experiment results in Section~\ref{sec:result} and detailed investigation of various parameters of the magnetic-enhanced MTS in Section~\ref{sec:discussion}. Finally, we draw our conclusion.

\section{Theory}
\label{sec:theory}

\begin{figure}%[tbh]
\centering
\includegraphics[width=\columnwidth]{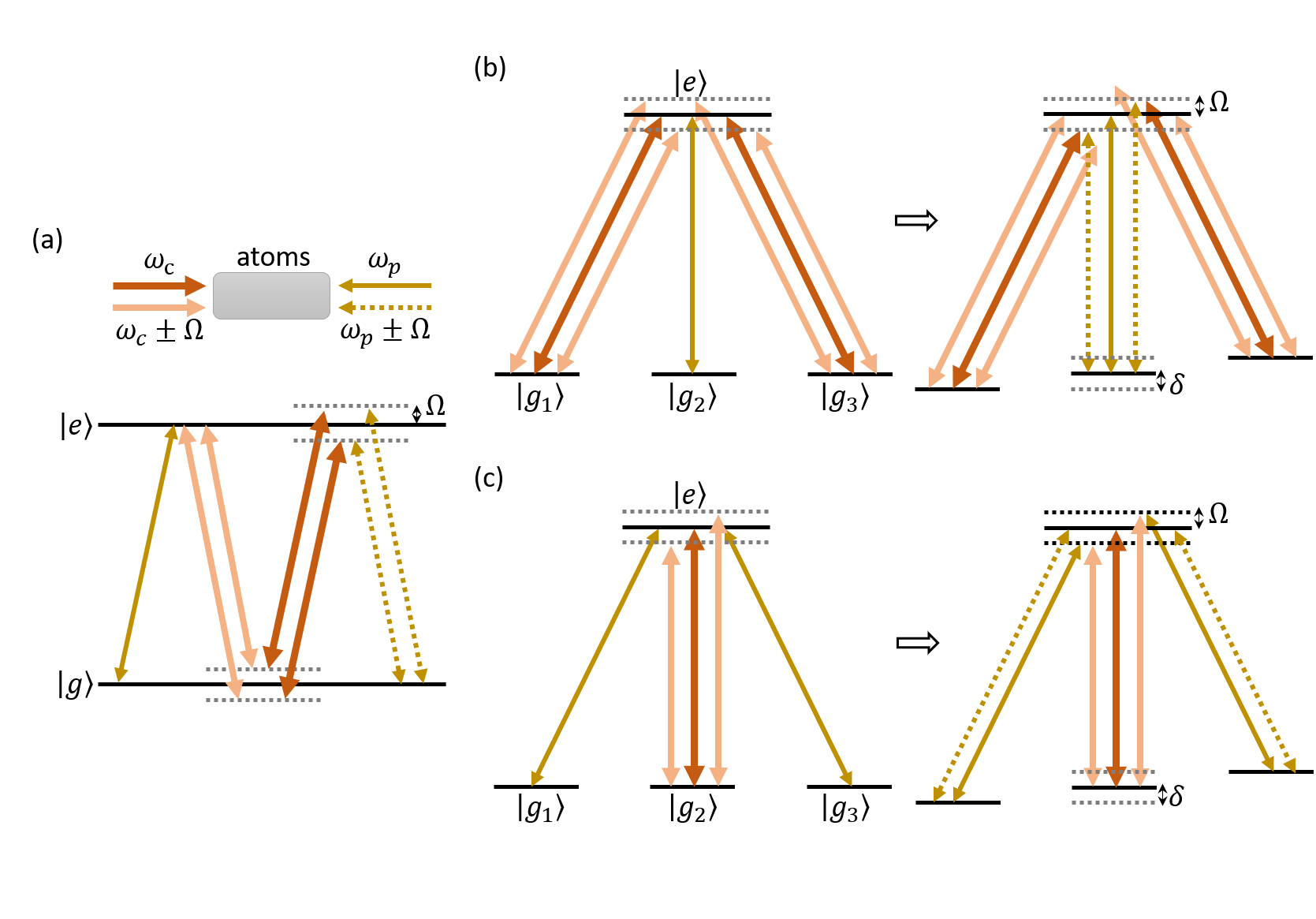}
\caption{Energy level diagrams for the two-level MTS (a), and the tripod-level MTSs of two orthogonal linearly-polarized configurations corresponding to the $^{87}\text{Rb}$ $D_2$ line transition $F_g\!=\!1\!\to\!F_e\!=\!0$ (b, c). Left and right of (b, c) are considering the degenerate and non-degenerate ground states respectively. $\left|g_{1,2,3}\right>$ and $\left|e\right>$, the ground and excited states;  $\omega_{p(c)}$, the probe (pump) beam, $\omega_c\!=\!\omega_p\!=\!\omega$; $\Omega$, the modulation frequency; $\omega_{c(p)}\!\pm\!\Omega$, the 1st-order sidebands of the pump/probe beam; and $\delta$, energy splitting of the adjacent ground states due to a bias magnetic field shown in Fig.~\ref{fig:ExpSetup}.}
\label{fig:theory}
\end{figure}

As shown in Fig.~\ref{fig:theory}(a), considering a two-level system, a pump beam ($\omega_c$) and a probe beam ($\omega_p$) are counter-propagating through the atoms. We assume the two beams have the same frequency $\omega_c\!=\!\omega_p\!=\!\omega$, and approximately equal amplitude ($E_c\!=\!E_p\!=\!E_0$). The modulation is applied on the pump beam with the frequency $\Omega$ to generate the sidebands by an acousto-optical or electro-optical modulator. The phase-modulated pump light can be expressed as
\begin{equation}
\begin{split}
E_c&=E_0\sin\left(\left(\omega_0\!+\!\Delta\right) t+\beta\sin\Omega t\right)\\
&=E_0\sum_{n=-\infty}^\infty J_n\left(\beta\right)\sin\left(\omega_0\!+\!\Delta\!+\!n\Omega\right) t,
\end{split}
\end{equation}
where $\omega\!=\!\omega_0\!+\!\Delta$, $\Delta$ is the frequency detuning from the line center $\omega_0$, $\beta$ is the modulation index (the ratio of the modulation depth to the modulation frequency), and $J_n$ is the $n$th-order Bessel function.

The modulation is transferred to the probe beam by the FWM effect, where the phase conjugated sidebands of the probe beam are generated. The strongest signal is always observed for the cycling transition where atoms cannot relax into any other states~\cite{Li2011}. Beat signal of the probe field and its sidebands is of the form~\cite{Jaatinen1995}
\begin{equation}
\begin{split}
S\left(\Omega\right)=& \frac{C}{\left[\Gamma^2+\Omega^2\right]^{1/2}}\sum_{n=-\infty}^\infty J_n\left(\beta\right)J_{n-1}\left(\beta\right)\\
& [\left(L_{(n+1)/2}+L_{(n-2)/2}\right)\cos\left(\Omega t+\phi\right)\\
& +\left(D_{(n+1)/2}-D_{(n-2)/2}\right)\sin\left(\Omega t+\phi\right)],
\end{split}
\label{eq:beat}
\end{equation}
where $C$ is a parameter that depends on the properties of the atomic transition, $\Gamma$ is the natural linewidth of the excited state, $\phi$ is the relative phase with respect to the modulation applied to the pump beam, and $L_n\!=\!{\Gamma^2}/{[\Gamma^2\!+\!(\Delta\!-\!n\Omega)^2]}$, $D_n\!=\!{\Gamma(\Delta\!-\!n\Omega)}/{[\Gamma^2\!+\!(\Delta\!-\!n\Omega)^2]}$.

In Fig.~\ref{fig:theory}, we only plot the first-order sidebands of the beams ($\omega_{c,p}\!\pm\!\Omega$). Figure~\ref{fig:theory}(a) presents two possible FWM transitions. The higher order harmonics are not shown, which can be ignored if $\beta\!<\!1$. Only considering the first-order pump sidebands, Eq.~(\ref{eq:beat}) can be simplified as
\begin{equation}
\begin{split}
S\left(\Omega\right)=& \frac{C}{\left[\Gamma^2+\Omega^2\right]^{1/2}} J_0\left(\beta\right)J_1\left(\beta\right)\\
& [\left(-L_{-1}+L_{-1/2}-L_{1/2}+L_1\right)\cos\left(\Omega t+\phi\right)\\
& +\left(D_{-1}-D_{-1/2}-D_{1/2}+D_{1}\right)\sin\left(\Omega t+\phi\right)].
\end{split}
\label{eq:beat01}
\end{equation}

In deep modulation regime ($\beta\!>\!1$), multiple sidebands are involved in the modulation transfer process, and the signal gradient and the peak-to-peak amplitude could be enhanced~\cite{Zhou2010}. The sine and cosine terms in the above equations describe the quadrature and in-phase components respectively. Furthermore, the beat signal can be simplified to
\begin{equation}
S\left(\Omega\right)=X\cos\left(\Omega t+\phi\right)\!+\!Y\sin\left(\Omega t+\phi\right).
\end{equation}

The maximum signal amplitude is obtained at $\tan(\phi)\!=\!Y/X$, that mixing the absorption and dispersion components~\cite{Jaatinen1995}. In our experiment, this phase angle $\phi$ can be accurately adjusted to obtain the optimal signal, as shown in Fig.~\ref{fig:ExpSetup}.

In the above discussion, we simply focus on a two-level system. Energy levels are complicated for real atoms, and the FWM process would involve more than two levels. The parameter $C$ in Eqs.~(\ref{eq:beat}) and~(\ref{eq:beat01}) is no longer a constant value. We need to consider all the possible transitions for calculating the MTS spectrum~\cite{Noh2011, Sun2016}. Here, instead of detailed theoretical calculation, a phenomenological analysis is presented for the Rubidium atoms in our experiment.

Considering Zeeman sub-levels of the $^{87}\text{Rb}$ $D_2$ line transition $F_g\!=\!1\!\to\!F_e\!=\!0$, it's a tripod-type energy structure. In Figs.~\ref{fig:theory}(b) and \ref{fig:theory}(c), we show two kinds of orthogonal polarization configurations, that (V, H) and (H, V) respectively. Vertical polarization (V) is parallel to the quantization axes, that defined by a bias magnetic field $\mathbf{B}$. And horizontal polarization (H) is a combination of the right- and left-circular polarizations, $\sigma^+$(R) and $\sigma^-$(L). The bias field splits the degenerate sub-level ground states $\left|g_{1,2,3}\right>$ by a frequency of $\delta\!=\!B\!\times\! 0.7\,\rm{MHz/Gauss}$, where $B$ is the magnetic strength. Due to Zeeman splitting $\delta$ between the adjacent magnetic substates, frequency shift $\Delta$ in Eq.~(\ref{eq:beat}) should be replaced with $\Delta\!-\!\delta/2$. Left and right of Figs.~\ref{fig:theory}(b) and \ref{fig:theory}(c) describe the degenerate and non-degenerate ground states respectively. The horizontal-polarized beam excites the transitions with $\Delta m\!=\!\pm 1$ where $\Delta m$ is difference of the magnetic quantum numbers between the excited and ground states, while the vertical-polarized beam excites the transitions with $\Delta m\!=\!0$. Without magnetic field, $\left|g_1\right>$ and $\left|g_3\right>$ are two-photon resonant via the excited state $\left|e\right>$, both for the applied field and the sidebands of it. As a result, the atoms are coherently trapped in a dark dressed state, known as the coherent population trapping (CPT) phenomenon~\cite{Arimondo1996}. Light on the transition of $\left|g_2\right>$ and $\left|e\right>$ couldn't interact with them, and nonlinear atom-light interaction will be significantly diminished. The MTS signal on this transition is hard to be observed. However, when a bias magnetic field exists, two-photon resonance condition between the ground states is suppressed. The higher the bias magnetic strength, the weaker the two-photon phenomenon. Instead, the FWM process dominates and the generated sidebands of the probe beam are strong enough for detection. This is the key for realization of the magnetic-enhanced MTS scheme.

Particularly, when $\Omega\!=\!\left(2n\!+\!1\right)\delta$ and $n\!\in\!\mathbb{Z}$, two-photon resonance between the adjacent magnetic substates is satisfied for combinations of certain counter-propagating beams, e.g. the 1st-order sidebands of the pump beam and the probe beam if $n\!=\!0$. It will enhance the FWM process in the Lambda-type energy structure. The generated sidebands of the signal field become maximal under this condition. When $\Omega\!=\!2n\delta$, two-photon resonance between $\left|g_1\right>$ and $\left|g_3\right>$ induces dark dressed state again, and the MTS signal becomes weak. A dip of the signal amplitude occurs. This is experimentally identified in Section~\ref{sec:discussion}. Competition between these optical coherence processes is complicated especially for large modulation index $\beta$. Considering the specific MTS scheme, we can draw a conclusion that, the enhancement of the MTS signal depends on the satisfaction of the two photon resonance between the probe beam and the components of modulated pump beams.

Besides optical coherence connecting the ground and excited states, there also exists Zeeman coherence among the ground sub-levels. Larmor precession induces redistribution of the atomic magnetic states. The applied magnetic field is several Gauss in our experiment, and the precession frequency is of orders of magnitude smaller compared with the optical pumping effect. Physical mechanisms of the magnetic-enhanced MTS discussed above still work. Yet, for other polarization configurations besides that shown in Figs.~\ref{fig:theory}(b) and \ref{fig:theory}(c), optical transition between the excited state and the three ground states $\left|g_{1,2,3}\right>$ may not be simultaneously covered. Without Zeeman coherence, the atoms would finally relax into the non-interact state, and the signal sidebands cannot be effectively generated. If a magnetic field exists, the atoms can be pumped out of the non-interact state by procession of the atomic magnetic moment. Such population transfer enhances the MTS signal by some degree compared with the case that no magnetic field exists (see results in Fig.~\ref{fig:polarization}).

\section{Experiment setup}
\label{sec:setup}

\begin{figure}%[tbh]
\centering
\includegraphics[width=\columnwidth]{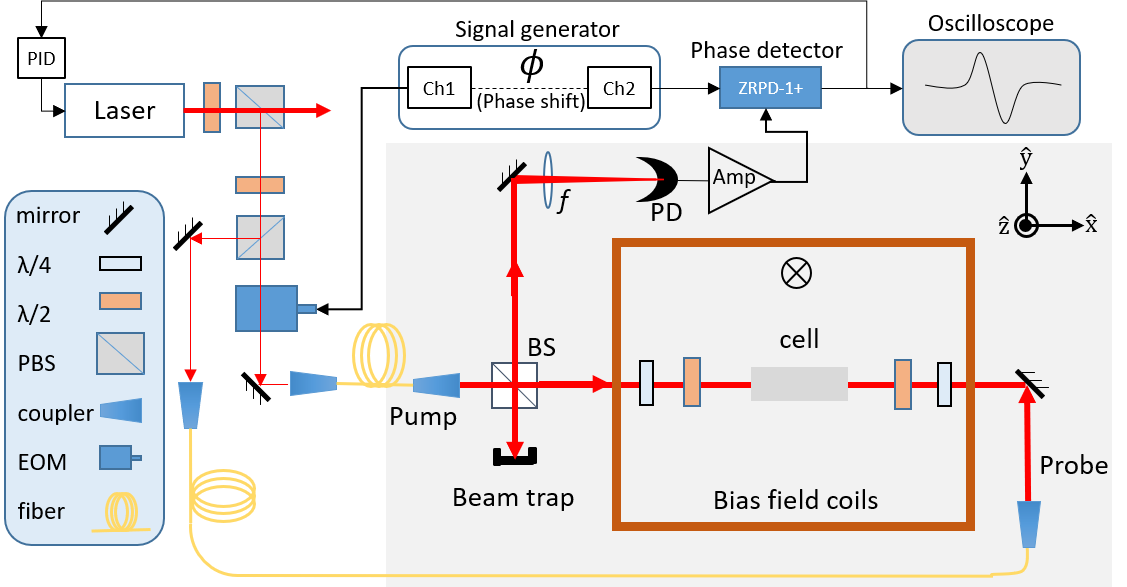}
\caption{Schematic for the magnetic-enhanced MTS and laser frequency stabilization. The bias field is along $z$-direction, perpendicular to the optical table. $\lambda/2$: half-wave plate, $\lambda/4$: quarter-wave plate, PBS: polarization beam splitter, coupler: laser collimator, EOM: electro-optic modulator, BS: 50/50\% beam splitter, $f$: lens, PD: photodiode detector, Amp: preamplifier, PID: proportion-integration-differentiation servo.}
\label{fig:ExpSetup}
\end{figure}

Experiment setup for the magnetic-enhanced MTS is illustrated in Fig.~\ref{fig:ExpSetup}. About 4\,mW power from a diode laser (Toptica DLpro @780\,nm) is picked off. It is separated into two laser beams by a $\lambda/2$ waveplate and a polarizing beam splitter (PBS) as the pump and probe beams respectively. The pump beam is phase modulated by an electro-optical modulator (EOM, Qu-big EO-Tx6L3-NIR) with a resonant frequency $\Omega$ in the range of 2 to 6\,MHz. The modulation signal is provided by one channel Ch1 of the signal generator 
(Agilent 335222). After coupled through single-mode fibers, the pump and probe beams go through the Rb vapor cell (length of 50\,mm) in a counter-propagating scheme. The two beams are expanded to about 3.5\,mm in diameter to enhance both the gradient and amplitude of the MTS signal~\cite{McCarron2008}, and overlap perfectly with each other to diminish the residual amplitude modulation~\cite{Jaatinen2009}. Polarization of the pump and probe beams can be tuned by adjusting the orientations of the waveplates ($\lambda/4$ and $\lambda/2$). The Earth and stray magnetic fields around the vapor cell are compensated bellow 10\,mGauss. A pair of square Helmholtz coils are placed to generate a bias magnetic field $\mathbf{B}$ along $z$-direction. The vapor cell is heated up to $44\,^{\circ} \text{C}$ to get a proper atom density for the spectroscopy.
\\

The probe beam and the accompanying sidebands are reflected by a beam splitter (BS) and focused on a photodiode detector (PD, Thorlabs PDA10A-EC). For orthogonal polarizations, the beam splitter can be replaced by a PBS. Reflection of the pump field on the BS is dumped by a beam trap. The photoelectric signal on the detector PD is amplified with a preamplifier (Mini-circuits ZFL-500+) and mixed with a signal from the channel Ch2 in a phase detector (Mini-circuits ZFPD-1+). The signal from Ch1 and Ch2 has the same frequency. The relative phase $\phi$ between them is adjustable. By tuning $\phi$, a dispersive error signal can be demodulated out of the phase detector and displayed on the oscilloscope. When the phase $\phi$ is adjusted to a proper value, the maximum peak-to-peak amplitude and zero-crossing gradient of a dispersive like error signal for the $D_2$-line transition $F_g\!=\!1\!\to\!F_e\!=\!0$ are obtained and recorded. For laser frequency locking, the error signal from the phase detector is delivered to a PID (proportion-integration-differentiation) controller and fed back to the laser diode.

In the experiment, the light gray part of the setup in Fig.~\ref{fig:ExpSetup} can also be replaced by an ordinary SAS/FMS setup, where the probe beam is dumped and the pump beam is reflected back through the vapor cell and recorded by the photodiode. The FMS signal is recorded. In the following section, comparison among various spectroscopy and laser stabilization methods is carried out to show the advantage of our magnetic-enhanced MTS method.

\section{Experiment result}
\label{sec:result}

\begin{figure}%[tbh]
\centering
\includegraphics[width=\columnwidth]{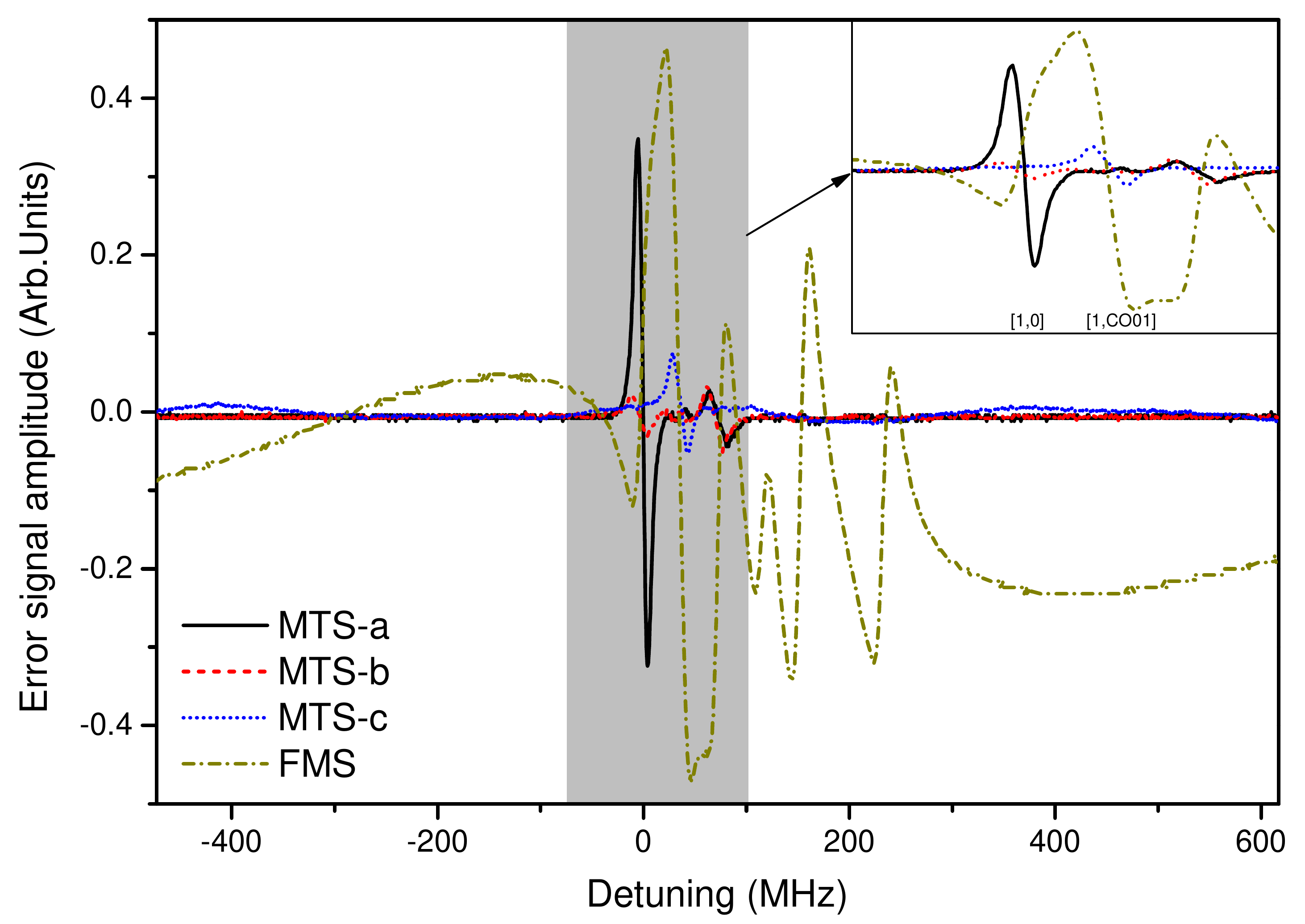}
\caption{Error signal comparison when the laser-diode frequency is scanned over the $^{87}\text{Rb}$ $D_2$-line transition of $F_g\!=\!1\!\to\! F_e\!=\!\{0,1,2\}$. Inset is enlarged of the gray part. MTS-a, the magnetic-enhanced MTS with orthogonal linear polarization setting; MTS-b, the ordinary MTS with orthogonal linear polarization setting; MTS-c, the ordinary MTS with parallel linear polarization setting; and FMS, the frequency modulation spectroscopy. The zero point in $x$-axis corresponds to the resonant transition of $F_g\!=\!1\!\to\!F_e\!=\!0$. CO01 is the crossover transition of $F_e\!=\!0\,\&\,1$. The modulation frequency is 4\,MHz. A 10-point moving average has been applied to all the data.}
\label{fig:CompTransition}
\end{figure}

\subsection{Error signal of the modulation transfer spectroscopy}

In Fig.~\ref{fig:CompTransition}, we compare the magnetic-enhanced MTS signal with the other spectroscopic methods. The modulation frequency of all these methods is 4\,MHz. The ordinary MTS means no bias magnetic field around the vapor cell. For the magnetic-enhanced MTS (MTS-a, black solid line), the bias field is about 1.5\,Gauss along $z$-direction (Fig.~\ref{fig:ExpSetup}). The probe beam is vertically polarized (V) along the field direction, while the pump beam is horizontally polarized (H). A maximal signal amplitude is obtained for the $^{87}\text{Rb}$ $D_2$-line transition of $F_g\!=\!1\!\to\!F_e\!=\!0$. The peak-to-peak amplitude is similar to that of the ordinary MTS on the cycling transition of $F_g\!=\!2\!\to\!F_e\!=\!3$ (not shown here). The beat signals of the other transitions nearby are effectively suppressed. By contrast, the red dashed line (MTS-b) is the ordinary MTS error signal around the transition of $F_g\!=\!1\!\to\! F_e\!=\!\{0,1,2\}$, with the same parameters of the MTS-a except that the bias field is off. Laser locking under such weak signal is almost impossible. To the best of our knowledge, there is no report of laser locking directly on this transition. Instead, our method enlarges the error signal more than twenty times and is suitable for laser frequency locking.

When changing polarization of the pump beam parallel with the probe beam, the MTS signal for the crossover transition of $F_g\!=\!1\!\to\!F_e\!=\!0\,\&\,1$ (MTS-c, blue dotted line) becomes obvious. Still, the signal amplitude is small compared with ours. It should be noted that the MTS signal for this parallel configuration is not a strict four-wave mixing process as discussed in Section~\ref{sec:theory}, but an incoherent process mediated by spontaneous emission~\cite{Park2013}. However, as the stray magnetic fluctuation would disturb the light transition, rigid magnetic shielding is demanding. The MTS signal is unstable for laser locking on this crossover transition.

We also present a frequency modulation spectroscopy around the transition of $F_g\!=\!1\!\to\!F_e$ (FMS, brown dashed-dotted line) in Fig.~\ref{fig:CompTransition}. The FMS signal is observed on a sloping background, approximating to the derivative of the Doppler-broadened absorption profile. Residual amplitude modulation leads to a DC offset drift of the error signal. And there are numbers of lines corresponding to those closely spaced hyperfine transitions. Laser frequency locking using this method cannot last for a long time and easily jumps to lock on unwanted frequency points.

\subsection{Stability of the laser lock}

Obviously, the magnetic-enhanced MTS signal for the transition of $F_g\!=\!1\!\to\!F_e\!=\!0$ shows considerable peak-to-peak amplitude with steep slope across the zero point, which is suitable for robust laser frequency stabilization. In this subsection, We investigate the laser frequency stability using our method, compared with the ordinary MTS and FMS methods.

\begin{figure*}%[tbh]
\centering
\subfigure{
\label{fig:locksignal}
\includegraphics[width=0.83\columnwidth]{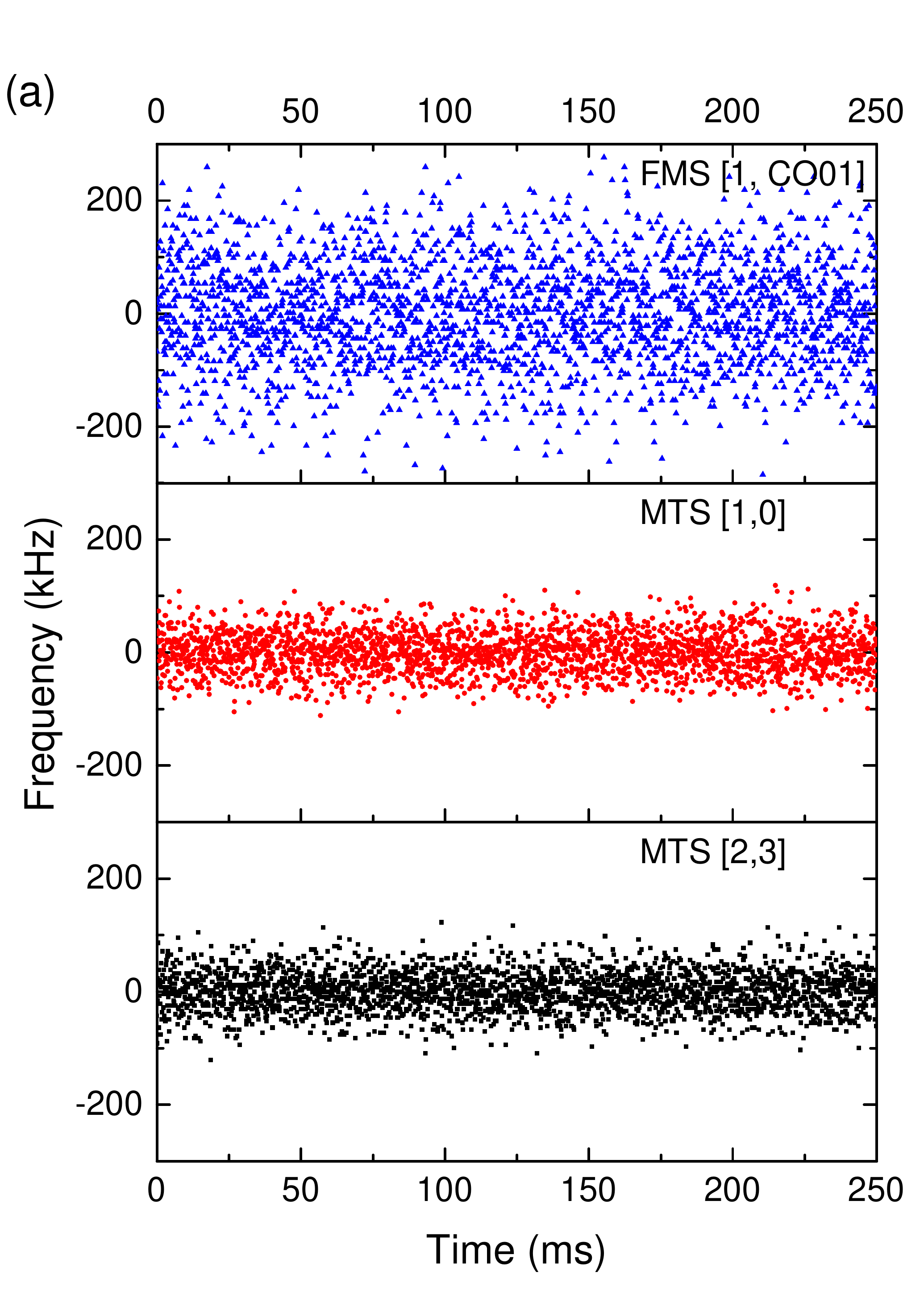}}
\hspace{0.0in}
\subfigure{
\label{fig:statis}
\includegraphics[width=\columnwidth]{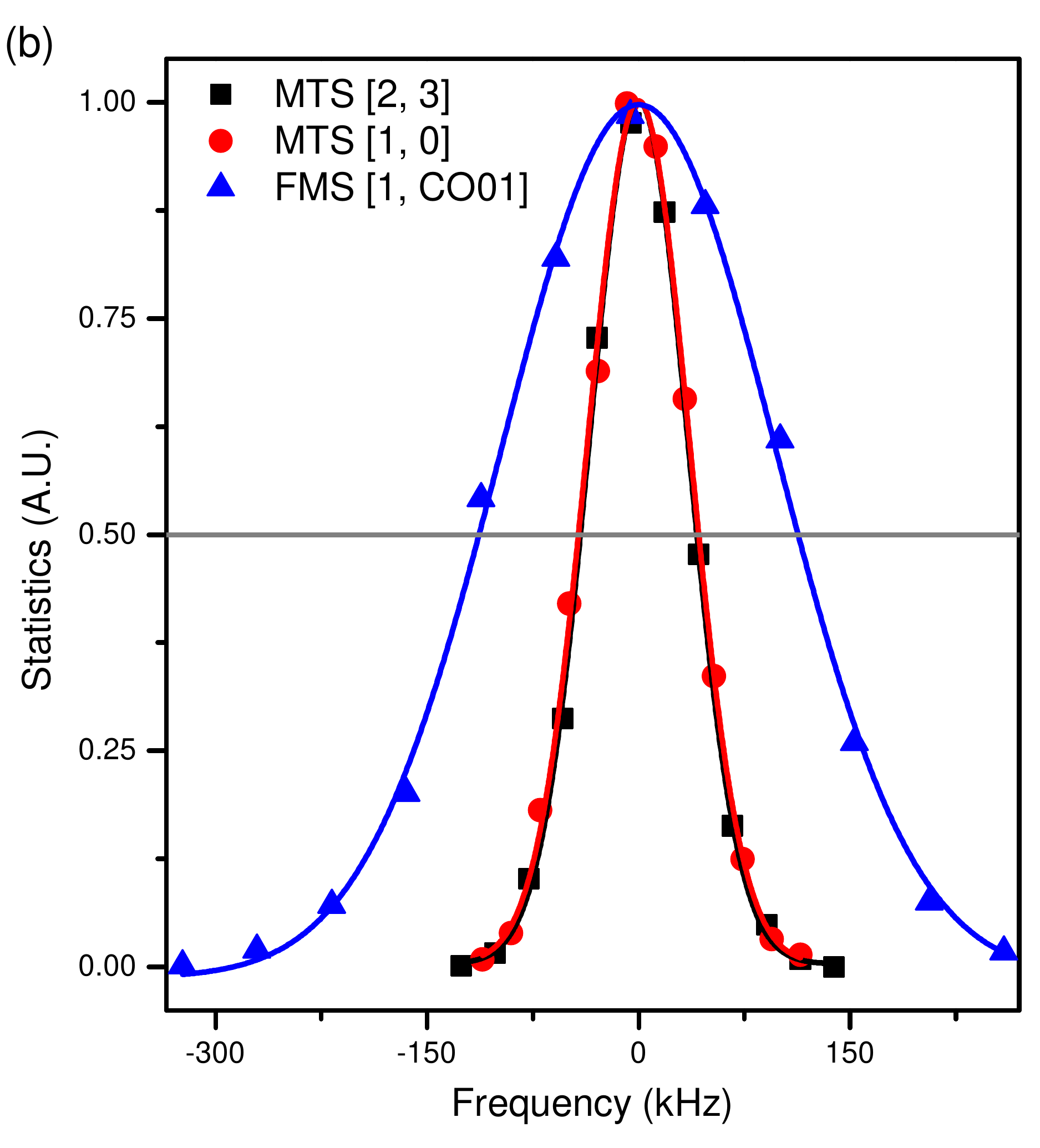}}

\caption{(a) Residual error signal fluctuation of three spectroscopy methods after laser lock; and (b) the corresponding statistical distribution of it. MTS [2, 3] (black square): the ordinary MTS lock on the transition of $F_g\!=\!2\!\to\!F_e\!=\!3$; MTS [1, 0] (red circle): the magnetic-enhanced MTS lock on the transition of $F_g\!=\!1\!\to\!F_e\!=\!0$; and FMS [1, CO01] (blue triangle): the FMS lock on the crossover transition of $F_g\!=\!1\!\to\!F_e\!=\!0\,\&\,1$. Sample interval in (a) is $100\,\mu s$ and the total duration is 0.25\,s; the $y$-axis in (b) is normalized to the peak value of the Gaussian fit. The FWHMs (full width at half maximum) of them are 82(1), 85(2) and 229(6)\,kHz respectively.}
\label{fig:CompErr}
\end{figure*}

{\it Error signal fluctuation after laser lock.} In Fig.~\ref{fig:CompErr}, we measure and compare the error signal fluctuations after locked by three spectroscopy methods, that locking with magnetic-enhanced MTS signal on the cycling transition of $F_g\!=\!1\!\to\!F_e\!=\!0$ (MTS [1, 0], red circle), with normal MTS signal on $F_g\!=\!2\!\to\!F_e\!=\!3$ (MTS [2, 3], black square), and with FMS signal on $F_g\!=\!1\!\to\!F_e\!=\!0\,\&\,1$ (FMS [1, CO01], blue triangle). The experiment parameters are the same with that mentioned in Fig.~\ref{fig:CompTransition}. This offers us a glimpse of the stability and noise estimation of laser locking. Figure~\ref{fig:locksignal} presents the residual signal fluctuations of the three spectroscopy after lock. Sample interval is 100\,$\mu s$, and record length is 0.25\,s. In the scanning mode, we calculate the frequency-to-time division of the $x$-axis according to the FMS. Then, we measure the signal slope at the locking point and convert the voltage of the beating signal into frequency. So, the $y$-axis is in unit of kilohertz. The MTSs provide more stable frequency lock compared with the FMS. This is because the four-wave mixing process in the MTS suppresses the background signal noise, while the error signal of the FMS suffers from a strong DC drift.

In Fig.~\ref{fig:statis}, we analyze statistical distribution of the error signal fluctuation. The $y$-axis is normalized to the peak value of the Gaussian fit. The full width at half maximum (FWHM) of the magnetic-enhanced MTS method (MTS [1, 0], red circle) is about $85\!\pm\!2$\,kHz, similar to that of the ordinary MTS locking on the transition of $F_g\!=\!2\!\to\!F_e\!=\!3$ ($82\!\pm\!1$\,kHz). Correspondingly, the FWHM of the FMS locking on $F_g\!=\!1\!\to\!F_e\!=\!0\,\&\,1$ is wide, about $229\!\pm\!6$\,kHz. The short-term stability of the FMS method is worse than that of the MTS methods. Frequency locking on this transition is quite unstable and would not be usually used in practice.

\begin{figure*}%[tbh]
\centering
\subfigure{
\label{fig:beatsignal}
\includegraphics[width=0.7\columnwidth]{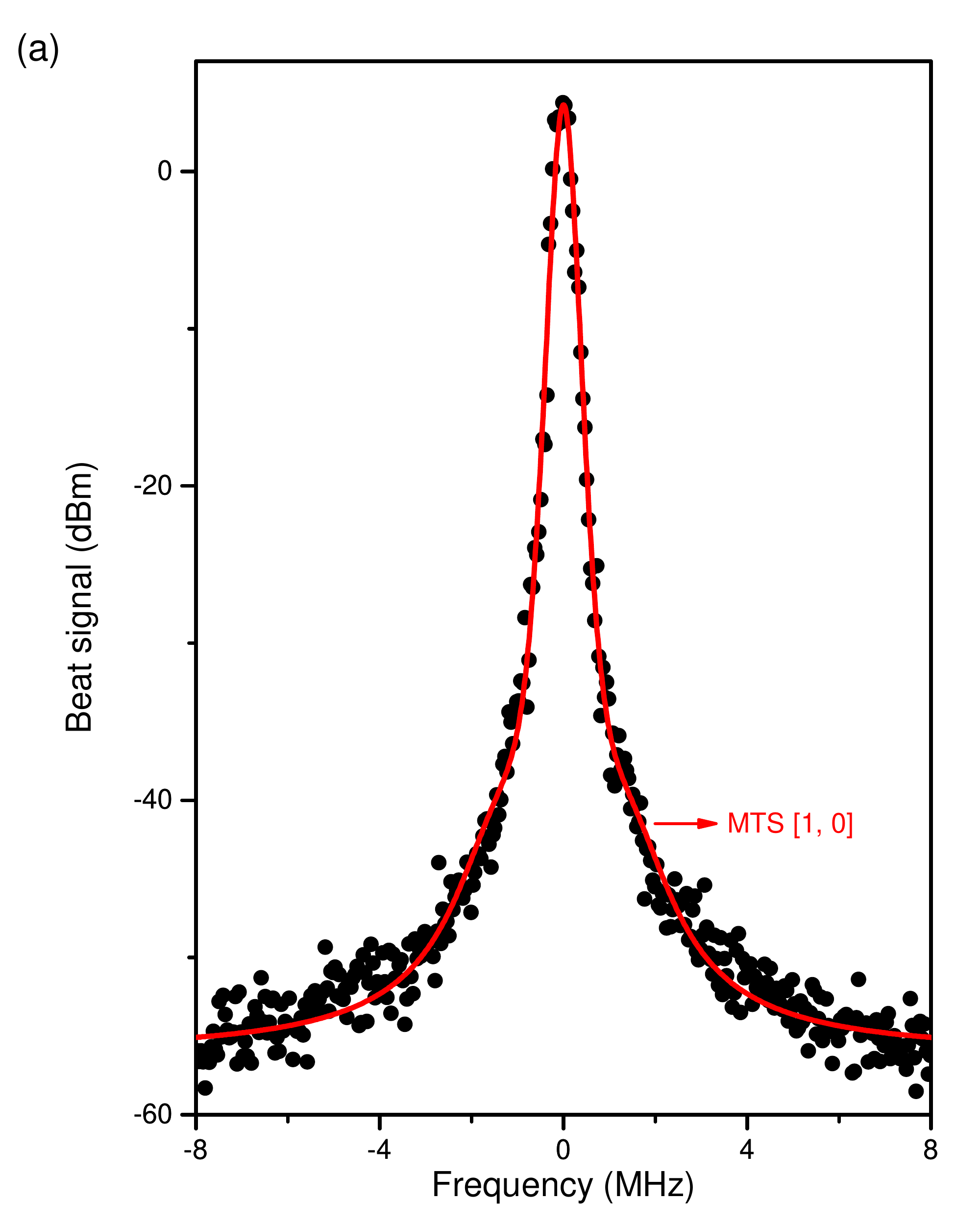}}
\hspace{0.0in}
\subfigure{
\label{fig:longterm}
\includegraphics[width=\columnwidth]{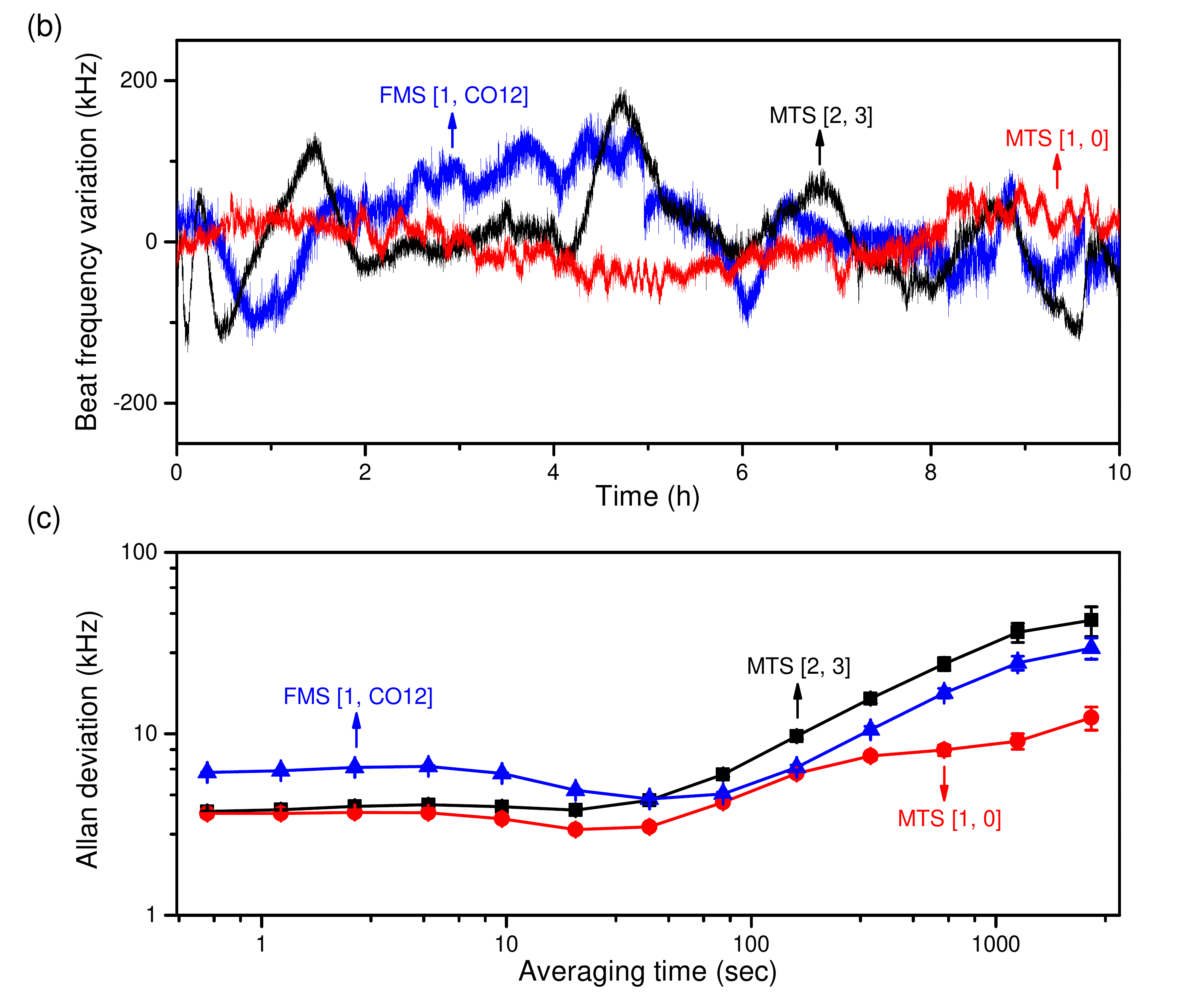}}

\caption{Measurement of laser linewidth and long-term stability by light beat of two independent laser diodes after lock. (a) Beat signal of the magnetic-enhanced MTS locking, fitting with a Lorentz-Gaussian function. Linewidth (FWHM) of the laser is about 209\,kHz. All the data is 10-times averaged. (b) Beat frequency variation of the two independent laser diodes after lock for 10 hours. Three spectroscopy methods as discussed in Fig.~\ref{fig:CompErr} are carried out here. The gate time of the frequency measurement is 0.5\,s. (c) Allan variance of the beat frequency.}
\label{fig:drift}
\end{figure*}

{\it Linewidth and long-term stability.} In the case with laser lock, the probe laser of our magnetic-enhanced MTS scheme is linear (V) polarized which induces the transition between the Zeeman sub-levels $m_{F_g}\!=\!0$ and $m_{F_e}\!=\!0$. Both of them are magnetic insensitive, which have zero 1st-order Zeeman shift. Considering the laser linewidth, the magnetic-induced frequency shift of the locking point is negligible. A promising long-term stability is expected. Here, two sets of identically frequency locked lasers with the frequency offset of 300\,MHz are applied to measure the laser linewidth and the long-term stability.
No $\mu$-metal magnet-shielding exists around the vapor cell. Three laser locking methods are carried out here, as discussed in Fig.~\ref{fig:CompErr}. Yet, laser locking using the FMS method is on the crossover transition of $F_g\!=\!1\!\to\!F_e\!=\!1\,\&\,2$, instead of $F_g\!=\!1\!\to\!F_e\!=\!0\,\&\,1$. That's because frequency drift of the latter is serious and unsuitable for practical laser frequency stabilization. Figure~\ref{fig:beatsignal} shows the beat frequency signal recorded by a spectrum analyzer (Agilent N9343C), where the lasers are locked using the magnetic-enhanced MTS method. Fitting with a Lorentz-Gaussian function, we obtain the laser linewidth of 209\,kHz by dividing the FWHM by $\sqrt{2}$. Linewidth using the other two methods are about 230\,kHz. It mainly depends on the original linewidth of the commercial external-cavity diode laser.

In Figs.~\ref{fig:drift}(b) and \ref{fig:drift}(c), we show the beat frequency variation for 10 hours and the Allan variance of it. A frequency counter (Agilent 53220) is used to measure the beat frequency with a gate time of 0.5\,s. When laser locking with the magnetic-enhanced MTS, the variation of the beat frequency has a standard deviation of 27\,kHz and a peak-to-peak drift of 162\,kHz. Assuming the frequency fluctuations are uncorrelated, standard deviation of the laser spectroscopy is 19\,kHz over 10\,h. The Allan deviation of the beat frequency indicates that the short-term stability is about 2.6\,kHz with an averaging time of 1\,s and the long-term stability is about 6.4\,kHz with an averaging time of 20\,min. For comparison, the beat notes are also measured when both spectroscopies work as the ordinary MTS (MTS [2, 3]) or FMS (FMS [1, CO12]) individually. For these two methods, standard deviation of the frequency variation is about 60\,kHz, three times larger than our method. Stability of the FMS spectroscopy is about 4.3\,kHz with an averaging time of 1\,s, while that of the ordinary MTS is the same with our method. For long-term stability, the Allan deviation of the magnetic-enhanced MTS is more than three times as good as the others. We owe the frequency instability of the ordinary MTS to the stray magnetic fluctuation and the FMS to the DC offset drift. Factors like drifts of the temperature and laser intensity also limit the long-term Allan deviation. The long-term stability can be further improved with more delicate magnetic shielding and laser intensity stabilization. Truly, our method is insensitive to the magnetic fluctuation and provides a more robust and accurate laser frequency locking compared with the other two methods.

\section{Discussion}
\label{sec:discussion}

In this section, we will show detail study of the magnetic-enhanced MTS, including various experiment parameters. Main factors that influence the magnetic-enhanced MTS signal include the bias magnetic strength, the light polarization configuration, and the modulation depth and frequency.

\begin{figure}%[tbh]
\centering
\subfigure{
\label{fig:polarization}
\includegraphics[width=0.95\columnwidth]{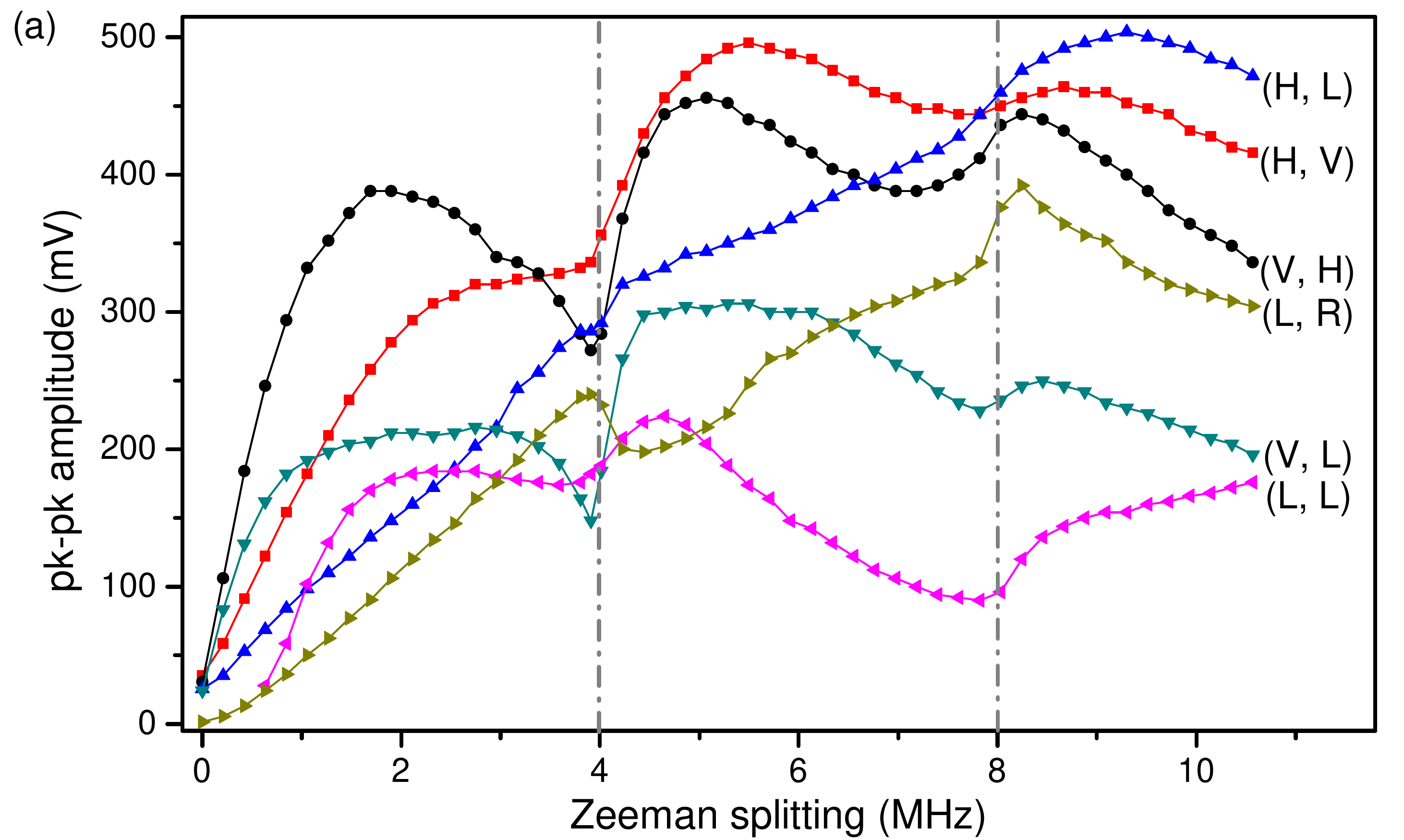}}
\hspace{0.0in}\\
\subfigure{
\label{fig:frequency}
\includegraphics[width=0.95\columnwidth]{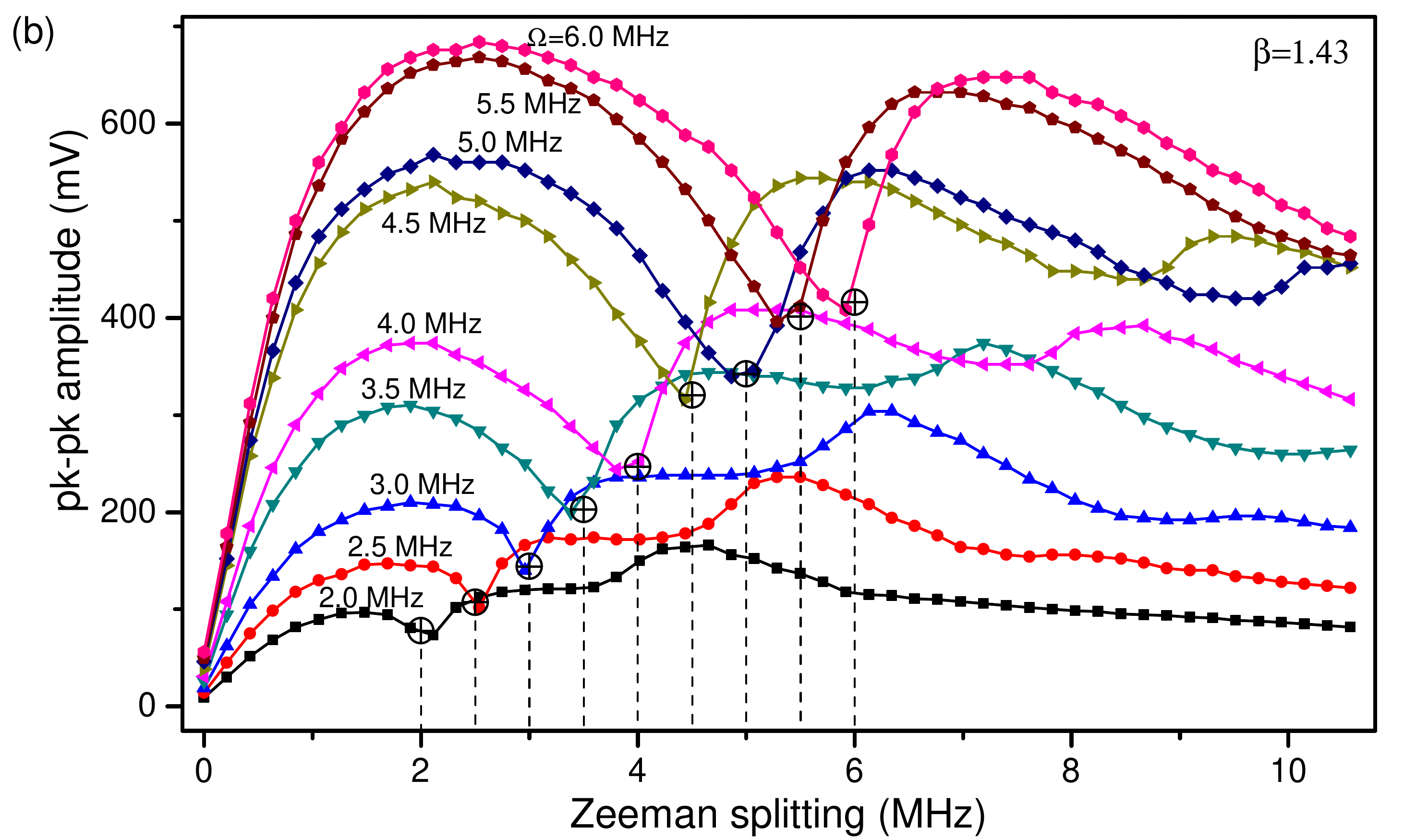}}
\hspace{0.0in}\\
\subfigure{
\label{fig:depth}
\includegraphics[width=0.95\columnwidth]{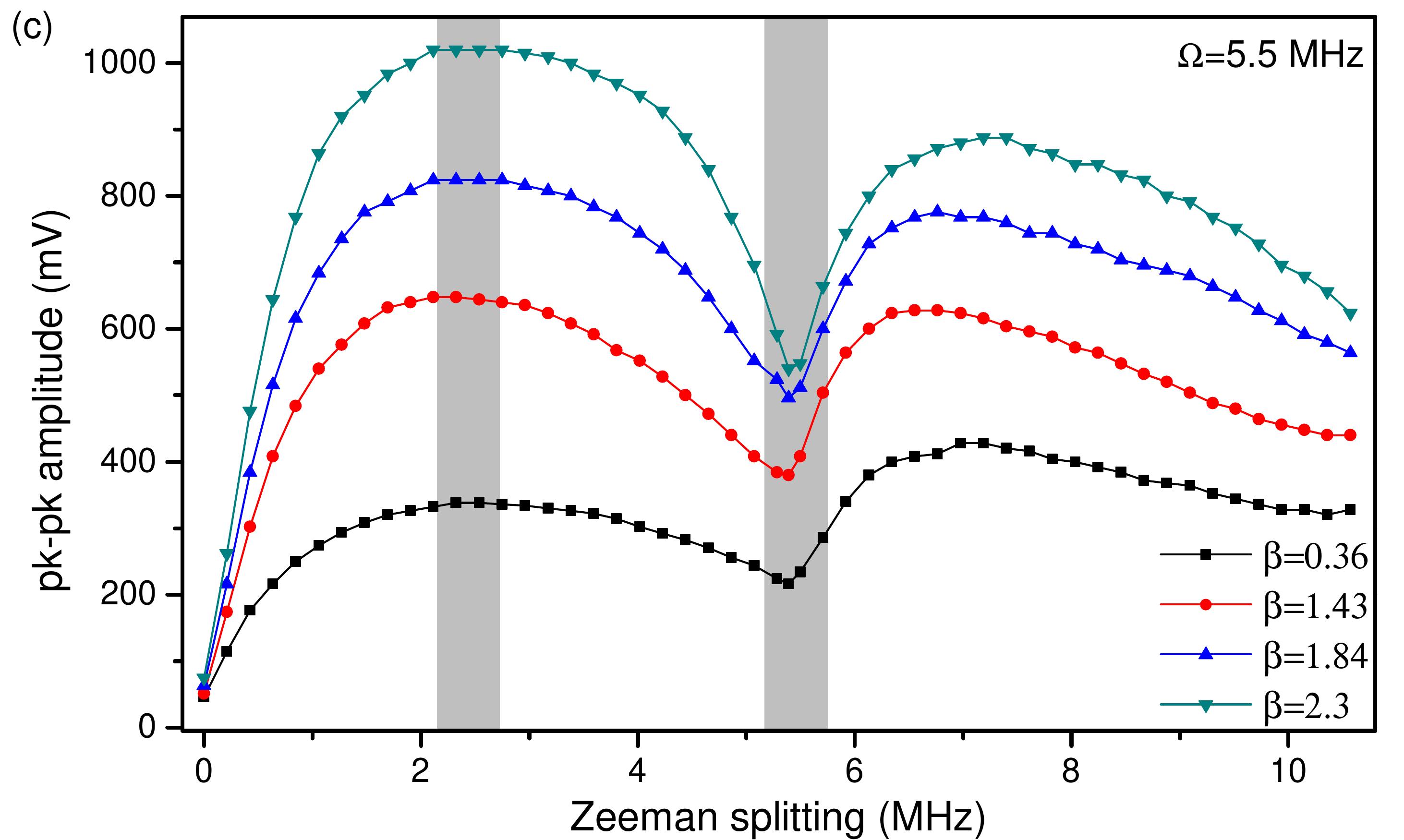}}
\caption{The peak-to-peak amplitude of the magnetic-enhanced MTS signal for various parameter settings, including (a) light polarization, (b) modulation frequency $\Omega$ and (c) modulation index $\beta$ when changing the bias magnetic strength $B$. Polarizations within parentheses in (a) corresponds to the probe and pump beams respectively. H/V, horizontal/vertical-linearly polarization; R/L, right/left-circularly polarization. Polarization configuration in (b) and (c) is (V, H). The $x$-axis is Zeeman splitting between the ground states of $F_g\!=\!1,m_{F_g}\!=\!\pm1$.}
\label{fig:MTSdetail}
\end{figure}

\subsection{Polarization}

Considering Zeeman sub-levels of the atoms, different polarization configuration of the pump and probe beams covers different energy structure. This has been discussed in Section~\ref{sec:theory}. In Fig.~\ref{fig:polarization}, we compare the peak-to-peak amplitude of the MTS signal for six type of polarization configurations on the transition of $F_g\!=\!1\!\to\!F_e\!=\!0$. The $x$-axis corresponds to Zeeman splitting between the magnetic sub-levels of $F_g\!=\!1,m_{F_g}\!=\!\pm1$. The modulation frequency $\Omega$ is set at 4\,MHz to obtain an optimal signal~\cite{McCarron2008}. Modulation index $\beta$ is about 1.25. Left and right labels of the polarization configurations are the polarization of the probe and pump beams respectively. The signals of linearly-polarized configurations are larger and increase faster with the bias field than those including of circular polarization. Obviously, the polarization combination (V, H) is a preferred choice. In this case, the probe light only excites the magnetic insensitive hyperfine transition $F_g\!=\!1,m_{F_g}\!=\!0\!\to\!F_e\!=\!0,m_{F_e}\!=\!0$, which provides a stable frequency reference for laser lock as proved in Fig.~\ref{fig:drift}. The first local maximum of the peak-to-peak amplitude at around $\delta\!=\!\Omega/4$ is chose for laser locking, as a comparatively small bias field is needed. Under this condition, two-photon detuning of all the optical transitions between $F_g\!=\!1,m_{F_g}\!=\!\pm1$ is the largest. The FWM effect becomes competitive. In our experiment, the magnetic strength is set to satisfy $\delta\!=\!\Omega/4$.

As shown in Fig.~\ref{fig:polarization}, there are packs or dips in the curves at around $\delta\!=\!\Omega/2$. This is because that two-photon resonance between the ground states $F_g\!=\!1, m_{F_g}\!=\!\pm 1$ is satisfied for high order sidebands of the laser beams. Such resonance would either diminish or enhance the nonlinear FWM process, depending on whether the two-photon resonant beams are counter-propagating or not. Roughly, the peak-to-peak signal amplitude is growing when increasing the bias magnetic strength. But for too large bias field, the amplitude decreases. According to the theoretical discussion in Section~\ref{sec:theory}, Zeeman coherence would transfer atoms out of the dark ground state where either no optical transition exists or the CPT condition is satisfied. It will benefit generation of the sidebands of the probe field and enhance the MTS signal. This is why the signal amplitudes are higher than the case of $\textbf{B}\!=\!0$. However, when Zeeman splitting is too large, sub-Doppler resonance condition may not be satisfied and the non-linear optical pumping becomes weak. Competition between these mechanisms induces what we have observed in Fig.~\ref{fig:polarization}.

\subsection{Modulation parameter}

In Figs.~\ref{fig:frequency} and \ref{fig:depth}, we choose the polarization configuration (V, H) of the probe and pump beams, and observe the influence of the modulation frequency and depth on the peak-to-peak amplitude respectively. Modulation index $\beta$ is about 1.43, and modulation frequency $\Omega$ is varying from 2\,MHz to 6\,MHz shown in Fig.~\ref{fig:frequency}. Here we can see more clearly that the dips appear at around $\delta\!=\!\Omega/2$. The peak-to-peak amplitude is increasing for large modulation frequency. When the bias magnetic field increases from zero, the MTS signal is initially enhanced, and reaches the maximum at the point $\delta\!=\!\Omega/4$. The signal amplitude decreases for strong magnetic field as sub-Doppler resonance is not satisfied.

In Fig.~\ref{fig:depth}, we investigates the influence of the modulation depth $\beta$. The modulation frequency is 5.5\,MHz. Large modulation depth increases the peak-to-peak amplitude of the MTS signals as the higher sidebands of the pump beam will generate multiple sidebands of the probe beam~\cite{Zhou2010}. However large modulation index also leads to serious fluctuation and noise of the signal. For laser frequency stabilization, we need a balanced consideration of the modulation frequency and depth.

\subsection{Zero-crossing gradient}

\begin{figure}%[tbh]
\centering
\includegraphics[width=\columnwidth]{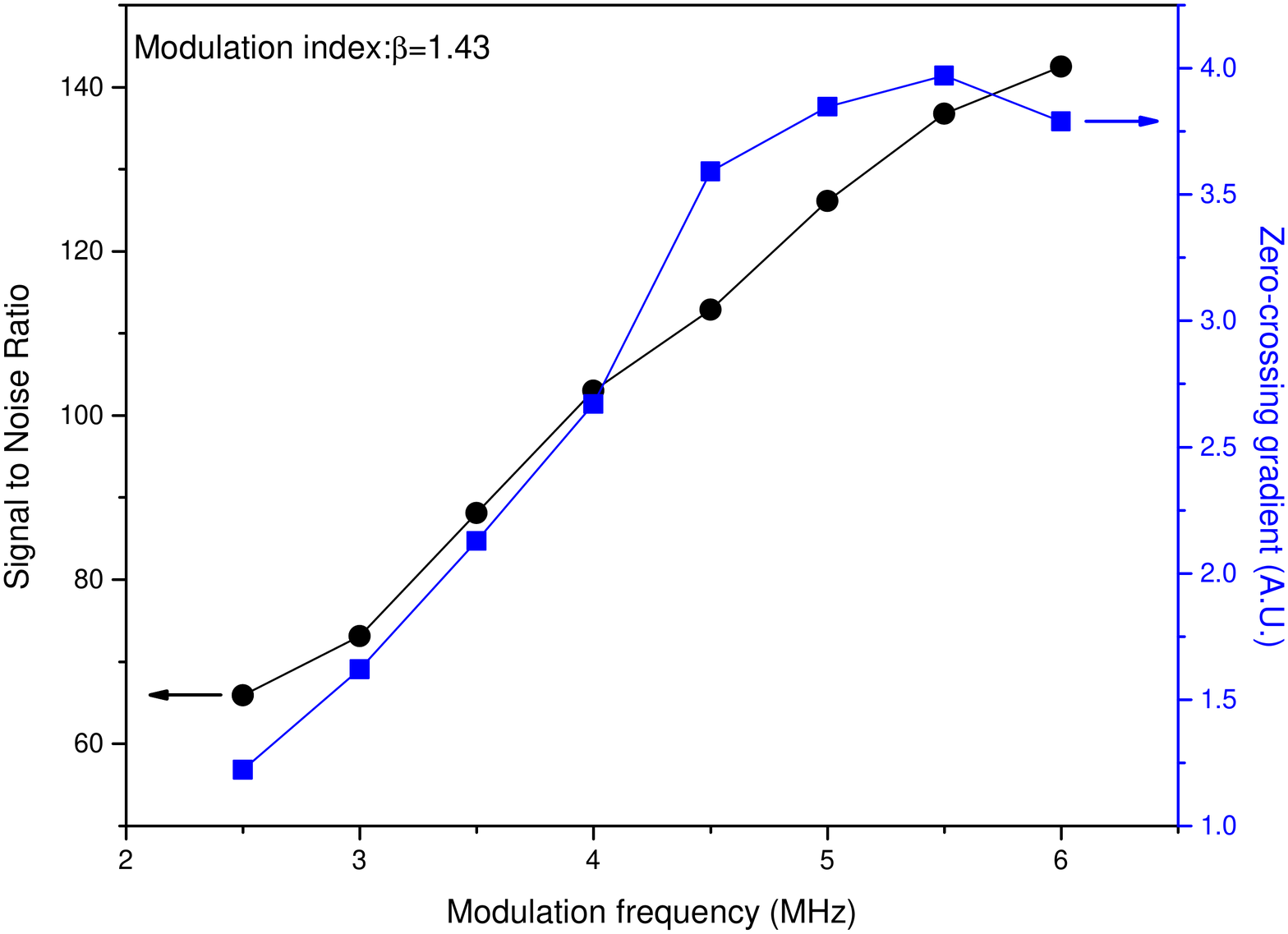}
\caption{The signal to noise ratio (left, black circle) and the zero-crossing gradient (right, blue square) of the MTS signal on the transition of $F_g\!=\!1\!\to\!F_e\!=\!0$ while changing the modulation frequency. Modulation index $\beta$ is fixed at 1.43.}
\label{fig:slop}
\end{figure}

Except the peak-to-peak amplitude, the zero-crossing gradient is also a key factor in laser locking. We will discuss it in this subsection. Optimum laser frequency stabilization is obtained when the MTS signal slope at the line center is a maximum and the background fluctuation noise is small. In Fig.~\ref{fig:slop}, we show the signal to noise ratio and zero-crossing gradient when changing the modulation frequency.
The zero-crossing gradient is the linear fitting slope of the spectroscopy at around the locking point when fixing the frequency scanning range. In the experiment, modulation index is 1.43. Polarization configuration of the probe and pump beams is (V, H). For each data point, the bias magnetic field $\textbf{B}$ and the phase $\phi$ between the probe and pump beams are optimized to obtain the maximum peak-to-peak amplitude.

According to Fig.~\ref{fig:slop}, the signal to noise ratio is growing with the modulation frequency. This is because the peak-to-peak amplitude increases with the modulation frequency while the fluctuation noise remains largely unchanged. In our experiment, the signal to noise ratio can be larger than 100:1. The zero-crossing gradient drops off for very high modulation frequency as the sub-Doppler condition is not satisfied. Peak value of the zero-crossing gradient at the line transition is about 5.5\,MHz, a little smaller than the natural linewidth $\Gamma$ ($\sim\!6\,\text{MHz}$). Thus in practice, modulation frequency close to $\Gamma$ is an optimal choice for laser stabilization.

\section{Conclusion}
\label{sec:conclusion}

In conclusion, we have experimentally investigated the magnetic-enhanced modulation transfer spectrum of $^{87}\text{Rb}$ $D_2$ line. By applying a bias magnetic field perpendicular to the propagating direction of the probe and pump beams, the magnetic-enhanced MTS signal on the transition of $F_g\!=\!1\!\to\!F_e\!=\!0$ is obtained. Both the slope and amplitude of the signal are significantly improved by optimal choice of the modulation parameters, like the polarization combination of the pump and probe lasers, the modulation depth, and the modulation frequency, etc. The signal to noise ratio of the magnetic-enhanced MTS signal can be larger than $100\!:\!1$. The locking point is immune to the magnetic fluctuation, and reduces the requirement of the magnetic shielding. Laser locking directly on this transition is achieved for the first time. It provides a robust and accurate laser frequency lock with excellent performance of long-term stability. In theory, physical mechanism of this method is also carefully analyzed. Our magnetic-enhanced MTS method may also work for other atoms or molecules, especially when no Zeeman splitting of the excited state ($F_{e}\!=\!0$) occurs, e.g. the Sodium atom. This technique provides an alternative choice for laser frequency stabilization and could be widely applied in the field of laser spectroscopy, laser cooling and trapping, and precision measurement with atoms and lasers.

\section*{Funding}
National Key R\&D Program of China (2016YFA0301601); National Natural Science Foundation of China (11604321,11674301); Anhui Initiative in Quantum Information Technologies (AHY120000); China Postdoctoral Science Foundation and the Chinese Academy of Sciences.

%%%%%%%%%% If using BibTeX:
%\bibliography{sample}

%%%%%%%%%% If preparing manually:

\end{document}